\documentstyle[epsfig]{mn}
\def\msun{M_{\odot}}
\def\simlt{\mathrel{\rlap{\lower 3pt\hbox{$\sim$}}\raise 2.0pt\hbox{$<$}}}
\def\simgt{\mathrel{\rlap{\lower 3pt\hbox{$\sim$}} \raise 2.0pt\hbox{$>$}}}
\def\lsim{\mathrel{\rlap{\lower 3pt\hbox{$\sim$}}\raise 2.0pt\hbox{$<$}}}
\def\gsim{\mathrel{\rlap{\lower 3pt\hbox{$\sim$}} \raise 2.0pt\hbox{$>$}}}

\def\Msun{M_{\odot}}

\begin{document}

\title[EM counterparts to MBH coalescences]{On the search of electromagnetic 
cosmological counterparts to coalescences of massive black hole binaries}

\author[Dotti, Salvaterra, Sesana, Colpi, Haardt]{M.~ Dotti$^1$, 
R.~Salvaterra$^1$, A.~Sesana$^1$, M.~Colpi$^2$, \& F.~Haardt$^1$\\
$1$ Dipartimento di Fisica e Matematica, Universit\`a  
dell'Insubria, Via Valleggio 11, 22100 Como, Italy\\
$2$ Dipertimento di Fisica G.~Occhialini, Universit\`a degli Studi di Milano
Bicocca, Piazza della Scienza 3, 20126 Milano, Italy}

\maketitle \vspace {7cm}

\begin{abstract}

We explore the nature of possible electromagnetic counterparts of 
coalescences of massive black hole binaries at cosmological distances 
detectable by the {\it Laser Interferometer Space Antenna} ({\it LISA}). 
An electromagnetic precursor, during the last year of
 gravitational wave (GW)--driven inspiral, or an afterglow within few
 years after coalescence, may highlight the position in the sky of 
galaxies hosting {\it LISA} sources. 
We show that observations of
 precursors and afterglows are mutually exclusive, depending on the 
 mass of the primary black hole.  Precursors are expected to occur in
 binaries where the primary (more massive) black hole is heavier than
 $\sim 10^7\;\Msun.$ They may correspond to on--off states of
 accretion, i.e., to a bright X--ray source decaying into quiescence
 before black hole coalescence, and are likely associated to
 disturbed galaxies showing signs of ongoing starbursts. 
Coalescences of lighter binaries, with masses $\simlt 5\times 10^6\;\Msun$, 
lack of any precursor, as gas is expected to be 
 consumed long before the GW-driven orbital decay. 
Such events would not be hosted by (massive)
 galaxies with an associated starburst, given the slow binary
 inspiral time compared to the typical time scale of starbursts. 
By contrast, coalescence, for such light
 binaries, is followed by an electromagnetic afterglow, i.e., an
 off--on accretion state rising in $\simlt 20$ yrs.  Using a cosmological
 merger tree algorithm, we show that future X--ray missions such as {\it XEUS} 
will be able to identify, in 20 yrs operation, 
almost all the massive BH binary detectable by {\it LISA}, and, in only 5 yrs, all the 
{\it LISA} sources at $z>6$.
\end{abstract}


\section{Introduction}

Massive black hole (BH) binaries are considered primary sources of
gravitational waves (GWs) detectable, throughout the
entire Universe (up to redshifts $z\gsim 10,$ Hughes 2002, Bender \& Hils 
2003), by the
space--based {\it Laser Interferometer Space Antenna} ({\it LISA};
Bender et al. 1994, Hils \& Bender 1995), designed to operate in the low--frequency
range ($3\times 10^{-5}-0.1$ Hz).  {\it LISA} will
observe the signal, emitted during the last year of inspiral and at 
coalescence, by BH binaries of characteristic
masses between  $10^4\msun/(1+z)$ and $10^9\msun/(1+z)$ (Haehnelt
1994; Jaffe \& Backer 2003; Wyithe \& Loeb 2003; Sesana et al. 2004,
2005; Alcubierre et al. 2005). 
In standard hierarchical cosmogonies,  a large number of BH binaries 
is predicted to form along the cosmic history, through collisions of (pre--)galactic structures 
(Begelman, Blandford, \& Rees 1980; Kauffmann \& Haehnelt 2000; Volonteri, Haardt, \& Madau 2003). 

{\it LISA} will operate as an all--sky monitor, and its data stream
will record the signals from a large number of sources
belonging to different populations of both galactic and cosmological
origin (Nelemans, Yungelson \& Portegies Zwart 2001; Benacquista et al. 2003;
Farmer \& Phinney 2003; Barak \& Cultler 2004).  
Given the difficulty in disentangling the different signals,
and the large number of noise sources, the degree of precision for the
determination of the source position on the sky is still an open issue
(Cutler 1998; Hughes 2002; Vecchio 2004; Kocsis et al. 2005).
Analysis of {\it LISA} data alone will provide  
a measure of a number of binary parameters not corrected for the source 
redshift (Vecchio 2004; Holtz \& Hughes 2005, Baker et al. 2006).
Thus, pinpointing the GW source to its electromagnetic (EM) counterpart 
within the {\it LISA} error cube would be of great importance. 
Although the detection of a GW signal from coalescing BHs   
would be an extraordinary event by itself,
the identification of an EM counterpart
would greatly increase the scientific payoff, allowing:  (1) to improve 
our understanding of the nature of the 
galaxy hosting coalescing massive BH binaries (e.g., galaxy type, colors, 
morphology, etc.), 
(2) to reconstruct the dynamics 
of the merging galaxies and of their BHs, and, most importantly, 
(3) to break the degeneracy
between the luminosity distance, which is directly measurable 
from the GW signal and the source redshift.
In this last case, {\it LISA} sources could be used
(1) as ``standard candles'' (``standard sirens'', Schutz 1986) to estimate 
fundamental cosmological parameters (Schutz 2002; Holz \& Hughes
2005), (2) to distinguish
between the measured redshifted BH masses 
and their rest-frame values, and   
(3), to improve our understanding of the accretion history of massive BHs
 comparing the EM luminosity with the BH  masses and spins.

Different possible EM counterparts of coalescing massive BHs have
been proposed in the last few years. 
Kocsis et al. (2005) proposed that  {\it LISA} events are associated to the brightest quasar in
the {\it LISA} error box, provided that the volume explored is limited 
to $z\simlt 1$ ($z\simlt 3$ in case of rapidly spinning BHs), guaranteeing an unambiguous identification. 
Armitage \& Natarajan (2002) suggested that a strong, potentially observable
accretion episode should anticipate the BH collision: 
gas trapped inside the binary orbit may produce a surface density spike, and a bright flare
with outflows just prior coalescence. 
Finally, Milosavljevic \& Phinney
(2005) argued that a near--Eddington  
X--ray afterglow should appear, delayed by few
years from the {\it LISA} detection. 
On such timescale, residual circumbinary gas accumulated 
during the viscous--migration of the less massive BH has time to refill
the accretion volume, and feed the central BH.

In this paper, we explore the proposed EM counterparts
to {\it LISA} events, relating 
their characteristics  to the physical properties of binary BHs
and of their host galaxies. We restrict our analysis to the BH binaries
with mass ratios $\gsim 0.01$, i.e. covering the expected range in 
hierarchical scenario of supermassive BH assembly (Sesana et al. 2005).
Although coalescences of binaries with more extreme mass ratio could be 
important sources of GW emission, the physical conditions for these events 
might be really different with respect to the cases here explored.
We address a number of questions:  
(1) What are the BH masses for which we can potentially
see an EM precursor or/and an EM afterglow?
(2) Does the host galaxy show visible signs of 
a collisional interaction (a  
starburst) during the last-year of BH inspiral?
(3) Will the next--generation of X-ray missions be able to identify 
{\it LISA} events at cosmological distances?  

The paper is organized as follows. In Sect. 2 we introduce characteristic 
radii and times associated to the circumbinary disc surrounding the BHs, and 
describe the latest stages of orbital decay, before GW--driven coalescence. 
In Section 3 we infer relations that are necessary (not sufficient) 
for the occurrence of an EM precursor phase, and explore the possibility
that tidal disruption of bound stars may trigger an episode of mass transfer 
leading to an X-ray flare. We then study the detectability
of EM afterglows with {\it XEUS}. 
In Section 4,  we discuss on the potential link between  
starburst--host--galaxies and  {\it LISA} events. In Section 5 we derive 
our conclusions with a sketch of the domains where EM preglows and 
afterglows can occur, in the BH mass parameter space.

\section{BH dynamics}
 
Coalescences of massive BHs are thought to be events associated with
mergers of (sub--)galactic structures at high redshifts, so that the 
orbital decay, from the large distance scale of a merger ($\sim 100$ kpc) 
to the subparsec--scale, likely occurs in a gas--rich, dissipative 
environment. In cosmological models of structure formation the 
typical mass ratio of close BH binaries observable by {\it LISA} is 
$\sim 0.1$, with a tail extending down to $\sim 0.01$ (Sesana et al. 2005). 
Thus, we will not consider here more extreme mass ratios.

Recently, Kazantzidis et al. (2005) explored
the effect of gaseous dissipation in mergers between gas--rich disc
galaxies with central BHs, using high resolution N--Body/SPH
simulations. The authors show how the presence of a cool gaseous component
is essential in order to bring, in minor mergers, the BHs to parsec scale 
separations. Gas infall deepens the potential well,
preserving the less massive galaxy against tidal disruption, and leading 
to formation of a close BH pair. \footnote{In collisionless mergers, the 
bulge of the lighter galaxy is tidally disrupted before the merger is
completed, leaving its central BH wandering in the outskirts of the 
remnant galaxy (Kazantzidis et al. 2005).}  
Moreover, the interplay between strong gas inflows and
star formation seems to lead naturally to the
formation, around the two BHs, of a massive circumnuclear gaseous
disc on scales of $\lsim 100$ pc, close to the numerical resolution limit.
Despite these advances, it is still difficult to establish the
dynamical properties of such self--gravitating discs at sub-parsec scales, as
well as to asses the characteristics of the BH orbits.\footnote{See
Mayer et al. 2006 for new preliminary simulations at higher resolution, 
which are designed to solve for the spatial and velocity structure of 
the circumnuclear discs and dynamics of the BH pairing process.}

Escala et al. (2005; see also Escala et al. 2004) and Dotti, Colpi, \&
Haardt (2006) have recently studied the dynamics of double
massive BHs (in the range $10^6\msun \leq M_{\rm BH} \leq 2.5\times
10^9\msun$) orbiting inside a circumnuclear disc modeled with a
Mestel profile. They found that orbital angular momentum losses via
dynamical friction against the gaseous background are sufficient to
pair the BHs into a bound Keplerian orbit (on distances of a few pc), and to
drive initially eccentric orbits into circular before the two BHs form 
a bound state (Dotti et al. 2006).
Escala et al. (2005) have also shown that gravitational torques excited by
ellipsoidal deformations of disc inner regions can bring the binary to even closer
distances. 
Torques can reduce the separation down to $\sim 0.1$ pc, the scale (for $M\simgt10^8\msun$) at which GW 
emission can bring the two BHs to the coalescence in less than an Hubble time.
In extrapolating the results of the highest resolution simulations
available to date, Escala et al. (2005) suggested that gas--driven
orbital decay leads the binary to coalescence in $\lsim 10^7$ yr.
All these simulations lack of enough spatial resolution 
to detail the transition boundary separating the
self--gravitating region of the disc from the region dominated by the
gravity of the binary.  
Inside this transition zone, the disc surrounding the binary is 
expected to be accreting through viscous stress. 
If the mass ratio $q$ ($q=M_2/M_1\leq 1$) of the binary BHs is $\ll1$, 
then we may estimate the
position of the radius $R_{\rm tr}$ at which such transition occurs, 
assuming that the disc is described by a Shakura--Sunyaev gas--pressure
dominated $\alpha$--disc (Shakura \& Sunyaev 1973), and that the disc mass
within $R_{\rm tr}$ is of order of the binary mass.

The disc mass in an $\alpha$--disc scales with radius as
\begin{equation}\label{eq:mdisc}
M_{\rm {disc}}(R)
\sim 3.5\times 10^8\;M_{\odot} \left(\frac{M_7}{\alpha}\right)^{4/5}
f_E^{3/5} R^{7/5},
\end{equation}
\noindent
where $f_E$ is the luminosity in Eddington units (computed using a radiative
efficiency $\epsilon=0.1$), and $M_7$ is
the reference mass of the central, more massive BH in units of $10^7\;\Msun$. 
Thus,
\begin{equation}
R_{\rm tr}
\sim 8 \times 10^{-2} 
\mbox{ pc } (1+q)^{0.71} M_7^{0.14} \alpha^{0.57} f_E^{-0.43}.
\end{equation}
Given our estimate of  $R_{\rm {tr}}$,  
one can fairly assume that dynamical friction can shrink the BH binary down 
to this separation.
Indeed, on the bases of known results of disc--planet interactions (see e.g. 
Papaloizou et al. 2006 and references therein) we expect
that $M_2$ perturbs gravitationally the surrounding 
viscous gas creating a low density, hollow 
region, called ``gap'', which separates the outer--circumbinary disc
from an inner  gas--poor region around the primary BH.
At $a_{\rm gap}$ the gravitational torque from the secondary BH
balances the gas--viscous torque.  

In the case considered here, the appearance of the gap would occur 
at a BH binary separation 
\begin{equation}
a_{\rm gap}\gsim a_{\rm tr}\sim 0.6 R_{\rm tr},
\end{equation}
\noindent
where the prefactor 0.6 is taken from the numerical simulations of Artymowicz 
\& Lubow (1994). The scale $a_{\rm tr}$ defines naturally the time for
binary coalescence set by the GW back--reaction, which corresponds  
to the longest time 
in absence of other processes of orbital angular momentum loss:
\begin{eqnarray}\label{eq:tagap}
t_{\rm GW}(a_{\rm tr})
&\sim& 2.5\times 10^{12}\;{\rm yr} \nonumber \\ 
&&\times M_7^{-2.43} \alpha^{2.29} f_E^{-1.71} q^{-1} (1+q)^{1.86} F(e)^{-1},
\end{eqnarray}
\noindent
where $F(e)$ takes into account the dependence of $t_{{\rm GW}}$
on the binary eccentricity $e$ 
(Peters 1964; following the result of Dotti et al. 2006 on BH
circularization, we consider $F(e)\sim 1$ hereafter).
Note that at $a_{\rm tr}$, $t_{\rm {GW}}\lsim 10$ Gyr 
for $M\gsim 3\times 10^7\;\Msun$ ($q=0.1$, $\alpha=0.1$, $f_E=1$). 
To accomplish coalescence on a shorter time,  
additional angular momentum losses are necessary. 

For $q\ll 1$, Armitage \& Natarajan (2002) have shown that, 
once the gap opens, gravitational torques exerted by the inner edge of the 
circumbinary disc can still shrink the orbit, so that the inspiral  
time is set by the disc--driven migration timescale. For their choice of 
binary and disc parameters, coalescence is reached in $\sim 10^7$ yr.
If we assume in our study that this process is efficient even for 
$q\lsim 0.1$, and that migration continues down to the  
binary separation at which the disc mass inside the orbit equals $M_2$ 
(Armitage private communication), i.e., 
\begin{eqnarray}
a_{\rm crit}&\sim& a_{\rm tr} [q/(1+q)]^{0.71} \nonumber \\ 
&\sim& 4.7\times 10^{-2} 
\mbox{ pc } q^{0.71} M_7^{0.14} \alpha^{0.57} f_E^{-0.43},
\end{eqnarray}
\noindent 
then the time needed to migrate from $a_{\rm gap}$ to $a_{\rm crit}$ 
would be 
\begin{equation}\label{eq:ttorque}
t_{\rm torque}\sim 4.2 \times 10^8 {\rm yr}\; 
\alpha^{-0.8} M_7^{-0.2} f_E^{0.4} 
(a_{\rm gap}^{1.4}-a_{\rm crit}^{1.4})
\end{equation}
\noindent 
(see eq. 1 of Armitage \& Natarajan 2002). 
From $a_{\rm crit}$ inwards, GWs can drive the binary to coalescence 
on a timescale  
\begin{equation}
t_{\rm GW}(a_{\rm crit})= t_{\rm GW}(a_{\rm tr}) (a_{\rm crit}/ a_{\rm tr})^4.
\end{equation}
\noindent
Note that now $t_{\rm GW}\lsim 10$ Gyr for $M_1\gsim 2\times 10^6\;\Msun$ 
($q=0.1$, $\alpha=0.1$, $f_E=1$). According to eqs. 6 and 7, the coalescence 
time of the binary BHs can be approximated as
\begin{equation}\label{eq:tfin}
t_{\rm coal} = {\rm min}\{t_{\rm GW}(a_{\rm tr}),
[t_{\rm torque}+t_{\rm GW}(a_{\rm crit})]\}.
\end{equation}

\section{Nuclear activity as EM counterpart}

\subsection{Circumbinary disc accretion: the precursor}

A galaxy showing nuclear activity can be a 
peculiar EM counterpart of a {\it LISA} event. In order to have an AGN 
precursor, part of the gas has to remain 
bound forming a small accretion disc at least around the more massive BH when 
the gap opens. The time of gas consumption, $t_{\rm duty}$, is given by  
\begin{equation}
t_{{\rm duty},1}=\frac{\epsilon}{1-\epsilon} \, \tau_{\rm Edd} 
f_E^{-1} \, \ln[1+(M_{{\rm disc},1}(a_{\rm gap})/M_1)],
\end{equation}
\noindent
where $\epsilon$ is the radiative 
efficiency, and $M_{{\rm disc},1}$ is the mass of the disc when the 
gap opens. $M_{{\rm disc},1}$ is obtained from eq. 1, using the estimate of the disc radius 
around $M_1$ as given by Artymowicz \& Lubow (1994), viz. $\simeq 0.45 a_{\rm gap}$ 
for our choice of parameters. 
The AGN is still in an ``on" state at binary coalescence only if 
\begin{equation}\label{eq:tcoal}
t_{\rm duty}(a_{\rm gap}) > t_{\rm coal}.  
\end{equation}

\begin{figure}
\begin{center}
\centerline{\psfig{figure=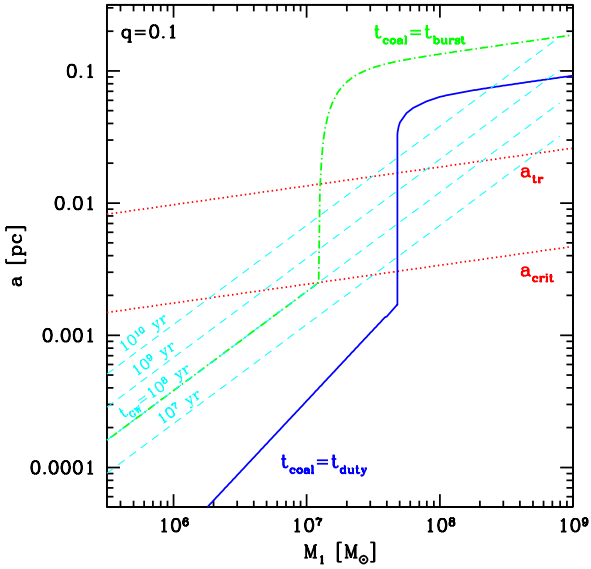,height=8cm}}
\caption{The curve labeled with $t_{\rm {coal}}=t_{\rm {duty}}$ shows, 
as a function of $M_1$, the largest BH binary separation for gap
opening in order to have nuclear activity at coalescence.  The curve
labeled with $t_{\rm {coal}}= t_{\rm {burst}}$ shows the separations
below which the BH binary merges on a time scale $t_{\rm
{burst}}=10^8$ yrs typical of a starburst (see discussion in Sect. 4).  
The smallest separation at
which the gap opens, $a_{tr}$, and $a_{crit}$ are shown with dotted
lines. Dashed lines correspond curves of constant coalescence time for
GW emission: $10^7$, $10^8$, $10^9$, and $10^{10}$ yr respectively,
from bottom to top. We assume here $q=0.1$, $\alpha=0.1$, $f_E=1$, and
$e=0$.}
\label{fig:q01}
\end{center}
\end{figure}

\begin{figure}
\begin{center}
\centerline{\psfig{figure=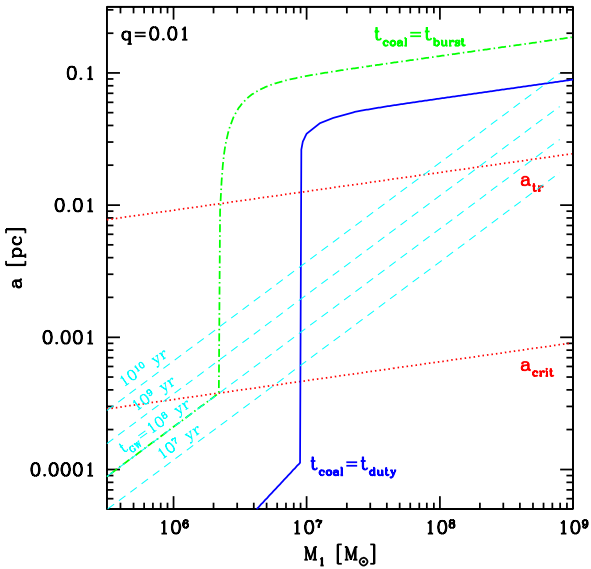,height=8cm}}
\caption{The same as Fig.~\ref{fig:q01} but for $q=0.01$.
}
\label{fig:q001}
\end{center}
\end{figure}

The presence of an AGN precursor varies with $M_1$ and $q$ as shown in Figure~\ref{fig:q01} (for $q=0.1$) and Figure \ref{fig:q001} 
(for $q=0.01$). The solid curve, labeled with $t_{\rm {coal}}=t_{\rm {duty}}$ 
in both figures, defines the critical BH binary separation below which 
gap opening still ensures nuclear activity at coalescence 
(assuming $\alpha=0.1$, $f_E=1$, and $e=0$). 
If the gap opens at larger separations, all the
gas in the accretion disc around $M_1$ is consumed before coalescence
(assuming negligible refilling of gas through the gap).
As reference, in the same figure we plot also $a_{\rm {tr}}$, 
to show the closest binary separation at which we expect the gap to open. 
So, for $M_1<1-5 \times 10^7\;\Msun$, assuming $q=0.01-0.1$, the gap
is opened well before $t_{\rm coal}=t_{\rm duty}$, implying that the 
accretion is completed long before coalescence. 
Larger masses, instead, imply that gas accretion is still going on. 
In this case, a bright X-ray counterpart could  be present in the error-box 
of {\it LISA}, given the large mass of the BHs.
During the last year of inspiral, and during the
BH plunge--in phase, gas dynamical perturbations and space--time curvature 
effects strongly affect the disc structure. 
Armitage \& Natarajan (2002) have shown that, if some 
residual gas is present around the  BHs just before  
coalescence, the accretion rate could be enhanced during the final GW 
driven inspiral. This process might trigger an episode of short-lived nuclear
activity, so a transient, highly variable on-off precursor can anticipate 
the BH merger, making easier the identification of the galaxy  
hosting the {\it LISA} event.

\subsection{Tidal disruption of bound stars}

A burst of nuclear activity might be excited by the tidal
disruption of a star bound to a BH, 
lasting  $\sim$ 1 yr (Rees 1988).
In order to produce an X--ray flare associated to a {\it LISA} event, 
the star has to be disrupted not earlier than one--year before the binary coalesces.

To asses the possible
occurrence of such disruption within one--year prior coalescence
two points must be satisfied: (a) the possibility for a star 
to survive in a orbit so bound that can interact with the secondary
BH one year before coalescence; (b) the 3--body interaction must
lead to the stellar tidal disruption instead of its ejection 
because of slingshot mechanism.

\begin{figure}
\begin{center}
\centerline{\psfig{figure=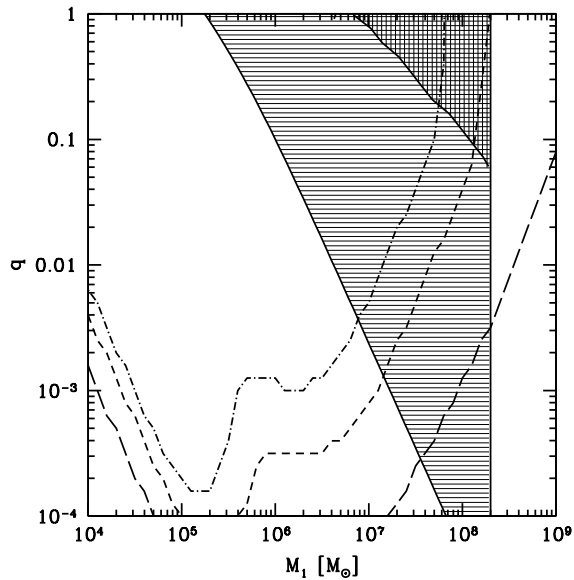,height=8cm}}
\caption{In the plane $M_1-q$, the region above the {\it long dashed (short 
dashed, dotted--dashed) line} limits {\it LISA} potential targets with 
{\it SNR$>$5} at $z=1(3,5)$. The horizontally dashed area marks systems 
with $r_{\rm {td}}<a_{\rm {GW,BH}}$, the {\it vertically dashed} area 
marks systems with $a_{\rm {GW,BH}}>a_{\rm {GW,\star}}$ (see text 
for discussion).}
\label{fig:tidal}
\end{center}
\end{figure}

To select binaries matching point (a), we can
compare three length scales: (1) the tidal disruption radius of a solar
type star orbiting around $M_1$ $r_{\rm td}\sim 8.3\times
10^{-6}\,{\rm pc}\,M_{1,7}^{0.67};$ (2) the separation at which a 
massive BH binary coalescence occurs in 1 yr: $a_{\rm GW,BH}\sim
3.7\times10^{-5}\,{\rm pc}\, M_{1,7}^{0.75}[q
(1+q)]^{0.25}F(e)^{0.25}$;
(3) the separation at which a solar type
star orbiting around the primary BH decays because of GW emission in,
say, $10^7$ yr which is the expected (shortest) lifetime of
a BH binary in a galaxy nucleus: $a_{\rm GW,\star}\sim3.7\times10^{-5}\,{\rm pc}\,
M_{1,7}^{0.5}$. The  radius $a_{\rm GW,\star}$ gives the 
dimension of the empty--cusp around the primary BH. 
A star can be perturbed in its motion 
and  be disrupted during the last-year
of BH binary inspiral only if $r_{\rm td}<a_{\rm GW,BH}$. The region
in the $M_1-q$ plane for which this condition holds is shown in
fig. \ref{fig:tidal} as horizontally shaded area, and is bounded on the
right by the condition $r_{\rm {td}}=6GM_1/c^2$: 
Note that for $M_1>2\times10^8 \Msun$ a solar--type star has  
a tidal radius smaller than the radius of the last--stable
circular orbit 
so that it would be swallowed by the BH rather than being disrupted. 
In addition, the existence of stars
around the two BHs on the verge
of merging, requires the size of the empty--cusp to
be inside the BH binary separation at one--year
from coalescence, i.e., $a_{\rm GW,BH}>a_{\rm GW,\star}.$
This region in the 
$M_1-q$ plane is indicated 
with the vertically shaded area.
The superposition of the two areas defines the parameter space of binaries
that are allowed to host a star so bound to $M_1$ that feels the presence 
of $M_2$ during the last year of the binary inspiral.

The point (b) is investigated by means of three body scattering experiments.
A detailed description of the 3--body procedure can be found in Sesana, Haardt
\& Madau 2006. Here we briefly review those technical features 
which are of interest for the present application. The three masses are treated   
as point--like particles, and the tidal disruption radius 
considered for the two BHs. We integrate the nine coupled, second order, differential equations
of motion, assuming Newtonian gravity, and the mutual force
acting on the two BH induced by gravitational wave in emission in 
quadrupole approximation (Iwasawa, Funato \& Makino 2005).
We set the initial conditions as follows: we consider circular BH orbits, 
at initial separation  $a_{\rm GW,BH}$; bound stars are then drawn from an isotropic 
velocity distribution spherically symmetric around $M_1$, with the constrain that the distance from  
$M_1$ at periastron is $>r_{\rm td}$, and at apoastron is $<a_{\rm GW,BH}-
r_{\rm td,2}$, where $r_{\rm td,2}$ is the tidal disruption radius of $M_2$.
The integration is stopped as soon as one of the following conditions is met: 
(1)  the star is kicked out by the binary with positive total energy, 
(2) the star crosses $r_{\rm td}$ or $r_{\rm td,2}$.  
During the integration of the orbit, as the binary shrink because of GW emission,
$M_2$ perturbs the stellar orbit around $M_1$. Typically the 
star is tidally disrupted by $M_1$ or $M_2$ well before it gets a positive energy 
because of slingshot mechanism. 
In a sample of about 5000 simulated orbits, we recorded $\sim 4950$ disruption 
events, i.e. $\sim 99\%$ of the total. In 
other words, (almost) all binaries satisfying our condition (a) above, are also 
producing a tidal disruption event.

\subsection{Disc accretion: the afterglow in the era of next generation X--ray satellites}

Right after coalescence, the central relic BH is embedded in a hollow region
surrounded by a gaseous disc. The gap will be re--filled on a time scale comparable to 
the gas viscous timescale. Milosavljevic
\& Phinney (2005) have shown that the AGN turns on after a time 
$t_{\rm on}\sim 7 (1+z) (M_{BH}/10^6\;\Msun)^{1.32}$ yr, creating an
Eddington--limited X--ray source of luminosity $L\sim 10^{43.5} (M_{BH}/10^6\;\Msun)$ erg s$^{-1}$. 
Such a  turn--on of 
nuclear activity could be the clear signature of a very recent coalescence, 
and would allow for the unambiguous EM counterpart 
identification among other X--ray sources in the 
{\it LISA} error cube (see Hughes 2002, Vecchio 2004 and Kocsis et al. 
2005 for a discussion on the {\it LISA} spatial resolution). 
The next generation of X--ray missions (such as 
{\it XEUS}\footnote{www.rssd.esa.int/index.php?project=XEUS} and 
{\it Constellation--X}\footnote{constellation.gsfc.nasa.gov}), 
expected to be operating simultaneously to {\it LISA},
will be able to detect these objects up to $z\simeq 20$ (Milosavljevic \& 
Phinney 2005).


The afterglow delay $t_{\rm on}$  is a function of BH mass.  
and, as an example, only BHs lighter then 
$3\times 10^6\;\Msun$ can be observed during an operation time of $\sim 30$ yr.
To quantify the number of X--ray afterglows detectable during 
the operation lifetime of next generation satellites, we must rely on 
a model for the BH assembly in hierarchical models of 
structure formation. 
To this aim, we started with the results of Sesana et al. (2004, 2005), which calculate 
the rate of BH binary coalescences as a function of redshift and mass (see their fig.~1). 
The contribution of massive BH binaries to the {\it LISA} data stream was computed  
using an extended Press--Schechter merger tree code able to follow the assembly of massive BHs (Volonteri, Haardt \& Madau 
2003\footnote{Note that Volonteri et
al. (2003) assume a short coalescence time invoking stellar dynamical 
processes with a continuos supply of low angular momentum stars.})
The total number of coalescences that can be detected in a 3 years {\it LISA} 
mission assuming a signal to noise threshold for detection S/N=5 
is shown with a solid line in fig.~4, and turns out to be $\simeq 35$, with a 
peak of detection around $z\simeq 5$. 
Now, if we   
(i) assume that the disc spectrum is described by a thermal modified 
blackbody for a nearly maximal spinning BH, so that most of the luminosity is 
emitted at rest--frame energies $h\nu\sim (0.5-5)$ keV (Milosavljevic \& 
Phinney 2005); (ii) neglect photo--electric absorption in the X--ray observed band \footnote{This 
assumption is justified particularly for high redshift sources, since in this case
we sample the high energy tail of the spectrum.}, and (iii), take as fiducial value for future X--ray mission a 
[0.5--2] keV flux limit $\sim 10^{-18}$ erg s$^{-1}$ cm$^{-2}$, 
we can compute the number of {\it LISA} sources detectable by
{\it XEUS} in 1, 5, and 20 yrs operation time (fig. 4). 
Already after 1 yr from the {\it LISA} detection, almost 25\% of the 
coalescence can be identified as X--ray sources. This fraction increases up to 
60\% (80\%) if longer lifetime for {\it XEUS} are considered. It is 
interesting to note that almost all the coalescences observed by {\it LISA} 
at $z\gsim 6$ have an EM counterpart detectable by future X--ray missions 
within the first 5 yrs after the GW burst. At these redshifts, {\it LISA}
error cube will be crowded by a large number of X--ray sources, so the sudden turn--on 
of a new X--ray source can give a peculiar fingerprint of the coalescence event. 
We also verify that increasing the flux limit (i.e., increasing the BH mass 
threshold for the afterglow identification) of future X--ray missions by a 
factor of, say, 10, our predicted counts remain valid, as 
afterglows are expected to have fluxes typically exceeding $10^{-17}$ erg/s/cm$^2$ in the 
X--ray band.

As far as EM counterparts in other wavelengths are concerned,  
we can speculate that a wiggling radio jet could be observable during the 
last phases of spiral in of the BHs (see as a review Komossa 2006 and 
references therein) in particular when the precursor is present that guarantees
the feeding of the BHs over most of their inspiral. This particular feature 
could also persist for several years after  the {\it LISA} detection.

\begin{figure}
\begin{center}
\centerline{\psfig{figure=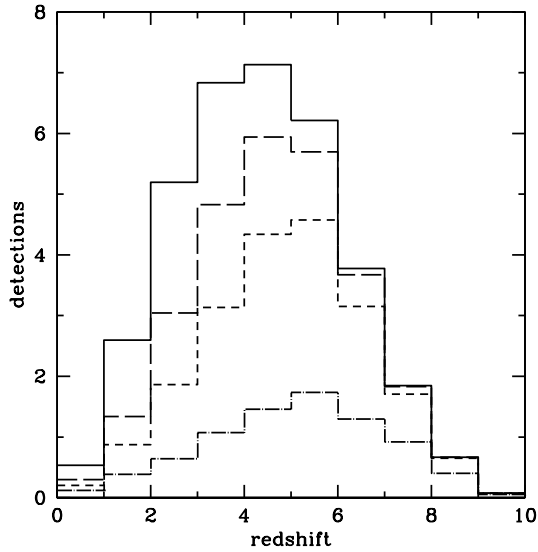,height=8cm}}
\caption{Number of coalescences as a function of redshift detectable by 
{\it LISA} in three years of operation (solid line; S/N $>5$)
and expected number of X--ray counterpart identifications by {\it XEUS} in 1, 
5, and 20 yrs with dashed--dotted, short dashed, and long dashed line, 
respectively (neglecting absorption). The standard $\Lambda$CDM cosmology
with $H_0=70$ km s$^{-1}$ Mpc$^{-1}$, $\Omega_m=0.3$, and $\Omega_\Lambda=0.7$
has been assumed.}
\label{fig:xeus}
\end{center}
\end{figure}

\section{Starburst galaxy as EM counterpart}

\begin{figure}
\begin{center}
\centerline{\psfig{figure=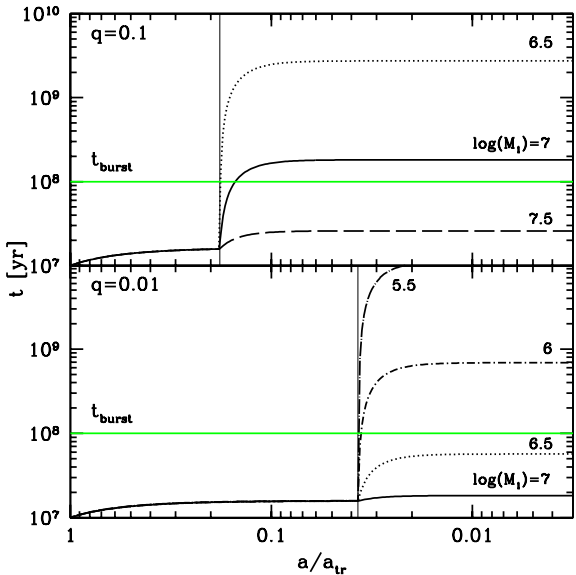,height=8cm}}
\caption{Run of $a/a_{tr}$ as function of evolution time. The vertical solid 
line marks $a=a_{crit}/a_{tr}$, while the horizontal line gives $t_{burst}$.
Different curves refer to different masses of the primary BH, labeled with
$\log(M_1/\Msun)$.
The top (bottom) panel shows the results for $q=0.1$ ($q=0.01$). We find that
the starburst is likely to be still active at the coalescence
only for $M_1>10\,(2)\times 10^6\;\Msun$.}
\label{fig:lisa}
\end{center}
\end{figure}

In standard hierarchical scenario of structure formation, the assembly of massive BH 
is connected to galactic mergers. Kazantzidis et al. (2005) have
performed N--body SPH simulations of binary equal-- and unequal--mass
(1:4) mergers of disc galaxies following the BH decay down to a
separation $\lsim 100$ pc, comparable to the numerical
resolution. These simulations show an intense burst of star formation 
excited when the two BHs form a pair. Almost 90\% of the central
gas is converted into stars in less than $t_{\rm
{burst}}\simeq 10^8$ yr, implying a star formation rate of 
$30-100\;\Msun$ yr$^{-1}$, comparable to that observed in  (ultra--)luminous infrared galaxies ((U--)LIRGs).  
The existence of such an intense burst of star formation associated with a 
merger event (see also Di Matteo, Springel \& Hernquist 2005; Springel, 
Di Matteo, \& Hernquist 2005), could be, in principle,
a peculiar feature of galaxies hosting coalescing BH binaries. 
The peculiar correspondence is observable if BH coalescence is attained
before starburst is completed, i.e., if BHs merge in less than
$10^8$ yr. As seen in Section~2, dynamical friction against stellar
and gaseous background is able to bring the BHs to the gap opening in
only $\sim 10^7$ yr.  As we already pointed out, for $q\lsim 0.1$, 
the interaction between the BHs and the
inner edge of the circumbinary disc can lead to coalescence on a
timescale $t_{\rm {coal}}$ given in Eq.~(\ref{eq:tfin}).

In Fig.~5 we show the run of $a/a_{\rm tr}$ with time $t$, assuming
that dynamical friction have driven the BHs to a separation $a_{\rm tr}$ (see eq. 3) 
in $\sim 10^7$ yr. The vertical solid line marks $a=a_{\rm {crit}}$ (eq. 5), the separation at which
viscous torques become inefficient.  For smaller separations, the
evolution of the binary is dominated by GW emission. Different curves
refer to different masses of the primary BH. The solid horizontal line
indicates the typical starburst duration, $t_{\rm {burst}}$.  The top
(bottom) panel shows the results for $q=0.1$ ($q=0.01$).  We find that
the starburst is likely to be still active at the coalescence
(i.e. $t_{\rm {coal}}\lsim t_{\rm {burst}}$) only for $M_1>10\,(2)
\times 10^6\;\Msun$.  On the contrary, for smaller masses, the
starburst is consumed before the BHs merge. In case dynamical friction
is inefficient until $a_{\rm {tr}}$, the starburst activity at
coalescence is still possible as shown in figs 1 and 2. The dot-dashed
line marks, in the $M_1-a$ plane, the locus $t_{\rm coal}=t_{\rm burst}$,  
assuming that at these separations the gap is already open, and that the
dynamics is dominated by viscous torques.  
Following this argument, the identification of a
(U-)LIRG in the {\it LISA} error cube could be a probable
EM counterpart of the GW event. 
Some caveat to this analysis has to be discussed:
(i) the more massive BHs ($\gsim 10^8 \; \Msun$) are preferentially hosted in 
gas poor early type galaxies that could be the outcome of a gas-rich major
merger. However there is the possibility that
dry mergers are important, i.e., mergers between gas poor
galaxies, and in this case an intense burst of star formation is not 
expected; (ii)  binaries with small $q$ may correspond to small
mass ratio of the interacting host galaxies, so that the associated 
starburst might be less powerful then what typically observed in  (U-)LIRG;
(iii) the presence of an intense starburst event in the last phases of galaxy
mergers could depend on the morphology of the two galaxies and in particular
on the relative size of the bulges, poorly addressed by numerical simulations
up to now. 
Nevertheless, the presence of a starburst galaxy in the {\it LISA} error cube
may indicate a preferential host of the BH coalescence.


\section{Summary and Conclusions}

\begin{figure}
\begin{center}
\centerline{\psfig{figure=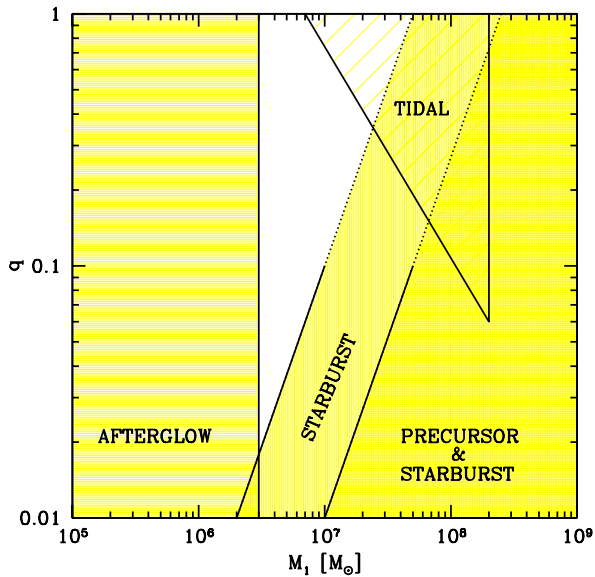,height=8cm}}
\caption{Summary sketch of the main results. Nuclear activity as a precursor 
is expected to occur in binaries with primaries heavier than 
$\sim 10^7\;\Msun$. 
Coalescences involving less massive BHs will lack of any precursor and, for 
$M_1\lsim 3\times 10^6\;\Msun$, are followed within 20 yr by an EM afterglow.
The small triangle labeled as ``tidal'' shows the region of the parameter space
where tidal disruption events of bound stars may be associated to the 
{\it LISA} detection. Finally, an intense starburst event may be still 
present at the coalescence of massive BHs ($M_1 \gsim 2 \times 10^6 \; \Msun$) 
embedded in gas rich galaxies.} 
\label{fig:conc}
\end{center}
\end{figure}

We have explored a number of possible EM counterparts of the coalescence of 
massive BH binaries at cosmological distances,  
assessing their detectability with future X--ray missions.
Our main findings are
summarized in fig.~\ref{fig:conc}.

EM precursor is expected to occur in binaries with primaries heavier
than $\sim 10^7\;\Msun$. If this event is related to a merger of
quasi equal mass, gas rich galaxies, star formation activity should be still
present at the coalescence time. In this case, the identification of a
starburst galaxy in the {\it LISA} error cube could be a probable
EM counterpart of the GW event.

AGN precursors, detectable during the last year of
inspiral, can be identified as variable X-ray sources switching to an 
off state. 
X--ray flares due to accretion of
matter pushed by the incoming secondary BH might be the fingerprint of
the {\it LISA} event. This on--off preglow would help in
identifying the host galaxy within the error cube. Note that the
extrapolation of our results to $q>0.1$ are only speculative, and
demands more detailed studies (solid lines ending with dots in fig.~\ref{fig:conc} 
indicate uncertain extrapolation).
EM afterglows associated to these massive binaries require onset
times in excess of $>100$ yr, and so exceedingly long expectation
times.  The triangle involving the heaviest BHs denotes the region
where a tidally disrupted star may be accreted in the last year before
coalescence, potentially causing a X--ray episode of accretion. This
would again appear as a preglow.

Smaller BHs ($M_1\lsim 3\times 10^6\;\Msun$) will lack of any precursor,
since we have found that the gas around the BHs is consumed long
before coalescence. Moreover, the starburst activity is also
unimportant, since the merging time of the BH binary is longer than
the typical starburst timescale. By contrast, a {\it LISA} event
involving light BHs is followed by an afterglow with rise--time less
than $20$ yr, i.e. the expected operation time of {\it XEUS}.  Using
cosmological merger tree algorithm, we have shown that {\it XEUS} 
will be able to identify almost all the predicted {\it LISA} events in
20 years of operation, and the totality of the events at $z>6$ in only 5 years.
There is a small transition region in between the light and heavy BHs
considered in this paper, where a {\it LISA} event may be only
associated to a starburst but shows no precursor.

The association of an EM counterpart to a {\it LISA} event
should be linked to a variable on--off (precursor) or off--on
(afterglow) state. In addition, we predict that the precursor is
related to mergers of heavy BHs, whereas afterglows are 
detectable only for lighter BHs. This dichotomy may be tested by
comparing the redshift of 
observed EM counterpart hosts with the redshifted masses
estimated analysing the {\it LISA} data stream (see, e.g., Vecchio 2004).

\section*{Acknowledgments}
We thank P.J.~Armitage, L.~Mayer, A.~Vecchio,  and M.~Volonteri for useful 
discussions.

\end{document}